\documentclass[prl,aps,showpacs,twocolumn,draft]{revtex4}
\begin{document}
\title{Nonlinear gauge transformation for a class\\ of Schr\"odinger
equations containing complex nonlinearities}
\author{G. Kaniadakis}\email{kaniadakis@polito.it}
\author{A.M. Scarfone}\email{scarfone@polito.it}
\affiliation{ Dipartimento di Fisica - Politecnico di Torino -
Corso Duca degli Abruzzi 24, 10129 Torino, Italy; \\
Istituto Nazionale di Fisica della Materia - Unit\'a del
 Politecnico di Torino, Corso Duca degli Abruzzi 24, 10129 Torino, Italy}
\date{\today}
\begin {abstract}
We consider a wide class of nonlinear canonical quantum systems
described by a one-particle Schr\"odinger equation containing a complex
nonlinearity. We introduce a nonlinear unitary transformation which
permits us to linearize the continuity equation.
In this way we are able to obtain a new quantum system obeying to a
nonlinear Schr\"odinger equation with a real nonlinearity. As an
application of this theory we consider a few already studied
Schr\"odinger equations as that containing the
nonlinearity introduced by the exclusion-inclusion principle,
the Doebner-Goldin equation and others.\\

\pacs{PACS numbers: 03.65.-w, 11.15.-q}
\end {abstract}
\maketitle
 Let us introduce the density of Lagrangian
\begin{equation}
{\cal L}=i\,\frac{\hbar}{2}\left(\psi^\ast\,\frac{\partial\,\psi}
{\partial\,t}-\psi\,\frac{\partial\,\psi^\ast}{\partial\,t}\right)-
\frac{\hbar^2}{2\,m}\,\Big|\frac{\partial\,\psi}{\partial\,x}\Big|^2-U[\psi,\,\psi^\ast]
 \ ,\label{lagra}
\end{equation}
which describes a class of one dimensional
nonrelativistic and canonical quantum systems.
In Eq. (\ref{lagra}) the nonlinear real potential $U[\psi,\,\psi^\ast]$ is a functional
of the fields $\psi$ and $\psi^\ast$.
We make use of the notation $U[a]$ to indicate that $U$ is a
functional of the field $a$ and of its spatial derivative, that is:
\begin{equation}
U[a]=U(a_0,\,a_1,\,a_2,a_3,\,\cdots) \ ,
\end{equation}
with $a_n=\partial^n\,a/\partial\,x^n$ and $a_0\equiv a$.\\
We will use also the hydrodynamic fields $\rho(x,\,t)$ and $S(x,\,t)$ defined as:
\begin{eqnarray}
&&\rho=\psi\,\psi^\ast \ ,\\
&&S=i\,\frac{\hbar}{2}\,\log\left(\frac{\psi^\ast}{\psi}\right) \ ,
\end{eqnarray}
which represent respectively the modulo and the phase of the field $\psi$:
\begin{equation}
\psi(x,\,t)=\rho^{1/2}(x,\,t)\,\exp\,\left[\frac{i}{\hbar}\,S(x,\,t)\right] \ .
\end{equation}
The evolution equations of the fields
$a\equiv\psi,\,\psi^\ast,\,\rho,\,S$ can be obtained from the
action of the system:
\begin{equation}
{\cal A}=\int{\cal L}\,dx\,dt \ ,
\end{equation}
by using the minimal action principle:
\begin{equation}
\frac{\delta\,{\cal A}}{\delta\,a}=0 \ .\label{act}
\end{equation}
Taking into account that the functional derivative of the potential term
in the action $U[a]$ is defined as \cite{Olver}:
\begin{equation}
\frac{\delta}{\delta\,a}\,\int U[a]\,dx\,dt=\sum_{n=0}\,(-1)^n\,
\frac{\partial^n}{\partial\,x^n}\,\left(\frac{\partial}
{\partial\,a_n}\,U[a]\right) \ ,
\end{equation}
it is easy to see that the evolution equation of the field $\psi$
can be obtained from (\ref{act}), posing $a\equiv\psi^\ast$ and is given by the
following Schr\"odinger equation which contains a complex nonlinearity:
\begin{equation}
i\,\hbar\,\frac{\partial\,\psi}{\partial\,t}=-\frac{\hbar^2}{2\,m}
\,\frac{\partial^2\,\psi}{\partial\,x^2}+W[\rho,\,S]\,\psi+i\,{\cal
W}[\rho,\,S]\,\psi \ .\label{sch}
\end{equation}
The real $W[\rho,\,S]$ and the imaginary ${\cal
W}[\rho,\,S]$ part are given by the following expressions:
\begin{eqnarray}
&&W[\rho,\,S]=\frac{\delta}{\delta\,\rho}\,\int U[\rho,\,S]\,dx\,dt \ ,\\
&&{\cal
W}[\rho,\,S]=\frac{\hbar}{2\,\rho}\frac{\delta}{\delta\,S}\,\int U[\rho,\,S]\,dx\,dt
\ .
\end{eqnarray}
The evolution equation for the field $\rho$ is obtained directly from
(\ref{act}), posing $a\equiv S$ (now $S$ is the field canonically
conjugated to the field $-\rho$). We obtain the equation:
\begin{equation}
\frac{\partial\,\rho}{\partial\,t}+\frac{\partial\,j_{_\psi}}{\partial\,x}=
\frac{\partial}{\partial\,S}\,U[\rho,\,S] \ ,\label{ec}
\end{equation}
where the quantum current $j_{_\psi}$ takes the expression:
\begin{equation}
j_{_{\psi}}=\frac{S_1}{m}\,\rho+\sum_{n=0}\,(-1)^n\,\frac{\partial^n}
{\partial\,x^n}\left(\frac{\partial}{\partial\,S_{n+1}}\,U[\rho,\,S]\right)
 \ .\label{cur1}
\end{equation}
We remark that Eq. (\ref{ec}) is the continuity equation and the term in
the right hand side represents a source for the field $\rho$. When
the conservation of the number of particles $N=\int\rho\,dx$
is required, the hypotesis that the potential $U[\rho,\,S]$ does not
depend on $S$ but only on its derivative must be introduced,
therefore the Eq. (\ref{ec}) takes the form:
\begin{equation}
\frac{\partial\,\rho}{\partial\,t}+\frac{\partial\,j_{_\psi}}{\partial\,x}=0
 \ ,\label{ec1}
\end{equation}
We note that Eq. (\ref{ec1}) can be obtained directly from
(\ref{sch}) and from its complex conjugate, performing the standard
procedure. We can see that the imaginary part ${\cal
W}(\rho,\,S)$ is responsible for the nonlinearity of the expression of the
current $j_{_\psi}$ (\ref{cur1}).\\
Let us introduce the following transformation for the field $\psi$:
\begin{equation}
\psi(x,\,t)\rightarrow
\phi(x,\,t)={\cal U}[\psi,\,\psi^\ast]\,\psi(x,\,t) \ ,\label{transf}
\end{equation}
which allows to eliminate the imaginary part of the evolution
equation of the field $\psi$, which corresponds also to
linearize the expression of the current $j_{_\psi}$.\\
The operator $\cal U$, generating this transformation, is unitary ${\cal U}^\dag={\cal
U}^{-1}$ and is defined by:
\begin{eqnarray}
\nonumber {\cal U}[\psi,\,\psi^\ast]&=&
\exp\left[i\,\frac{m}{\hbar}\,\sum_{n=0}\,(-1)^n\right.\\
&\times&\left.\int\frac{1}
{\rho}\frac{\partial^n}{\partial\,x^n}\left(\frac{\partial}{\partial\,S_{n+1}}\,
U[\rho,\,S]\right)\,dx\right] \ . \label{gauge}
\end{eqnarray}
If we write the field $\phi$ in terms of the hydrodynamic fields $\rho,\,\sigma$:
\begin{eqnarray}
\phi(x,\,t)=\rho^{1/2}(x,\,t)\,\exp\,\left[\frac{i}{\hbar}\,\sigma(x,\,t)\right] \ ,
\end{eqnarray}
and, due to the unitarity of the transformation, the modulo
of $\phi$ is equal to the modulo of the field $\psi$, while the
phase $\sigma$ is given by:
\begin{equation}
\sigma=S+\frac{m}{\hbar}\,\sum_{n=0}\,(-1)^n\,\int\frac{1}
{\rho}\frac{\partial^n}{\partial\,x^n}\left(\frac{\partial}{\partial\,S_{n+1}}\,
U[\rho,\,S]\right)\,dx \ .
\end{equation}
By accepting the statement made by Feynman and Hibbs (\cite{Feynman},
p.96): "{\sl Indeed all measurements of quantum-mechanical systems could be
made to reduce eventually to position and time measurements}", (see also
\cite{Doebner1}) the two wave functions $\psi$ and $\phi$ represent the same physical
system and, as a consequence, we can interpret the Eq. (\ref{transf}) as a
nonlinear gauge transformation of the function described by Eq.
(\ref{sch}).\\
From Eq. (\ref{sch}) and taking into
account the transformation (\ref{transf}), it is easy to obtain the following
evolution equation for the field $\phi$:
\begin{eqnarray}
\nonumber
&&i\,\hbar\,\frac{\partial\,\phi}{\partial\,t}=-\frac{\hbar^2}{2\,m}\,\frac{\partial^2\,\phi}
{\partial\,x^2}+W[\rho,\,S]\,\phi \\
\nonumber&&-\frac{1}{2}\,m\,\left[\sum_{n=0}\,(-1)^n\,\frac{1}{\rho}\,
\frac{\partial^n}{\partial\,x^n}\left(\frac{\partial\,U[\rho,\,S]}{\partial\,S_{n+1}}
\right)\right]^2\,\phi\\
\nonumber &&-m\,\sum_{n=0}\,(-1)^n\,\frac{\partial}{\partial\,t}
\left[\,\int\frac{1}{\rho}\,\frac{\partial^n}{\partial\,x^n}
\,\left(\frac{\partial\,U[\rho,\,S]}{\partial\,S_{n+1}}\right)\,dx\right]\,\phi\\
&&-\sum_{n=0}\,(-1)^n\,\frac{S_1}{\rho}\,\frac{\partial^n}{\partial\,x^n}\left(\frac{\partial\,U[\rho,\,S]}
{\partial\,S_{n+1}}\right)\,\phi \ .\label{schp}
\end{eqnarray}
Note that the nonlinearity appearing in Eq. (\ref{schp}) is now real.
The continuity equation of the system takes
the form:
\begin{equation}
\frac{\partial\,\rho}{\partial\,t}+\frac{\partial\,j_{_\phi}}{\partial\,x}=0
\end{equation}
where the current $j_{_{\phi}}$ has the standard expression of the
linear quantum mechanics:
\begin{equation}
j_{_{\phi}}=\frac{\sigma_1}{m}\,\rho \ .\label{cur2}
\end{equation}
The gauge transformation (\ref{transf}) and (\ref{gauge})
makes real the complex nonlinearity in the evolution equation,
and makes non canonical the new dynamical system.
However, this transformation may be useful to describe the evolution
of system by means of an equation containing a real nonlinearity. We
note that nonlinear transformations have been introduced and used
systematicaly for the first time in order to study nonlinear
Schr\"odinger equations as the Doebner-Goldin one in Ref. \cite{Doebner1}.\\
Let us apply the proposed transformation to a few already known equations
describing systems of collectively interacting particles.\\
The first example is given by the canonical Doebner-Goldin equation
($c_1=-c_4=1,\,c_2=c_3=c_5=0$) \cite{Doebner1,Doebner2}
that can be obtained from (\ref{lagra}) where the potential
$U[\rho,\,S]$ has the following form:
\begin{equation}
U[\rho,\,S]=\frac{D}{2}\,(\rho_1\,S_1-\rho\,S_2) \ .\label{fokker}
\end{equation}
A complex nonlinearity is generated in the evolution equation of the
field $\psi$, with real and imaginary part given respectively by:
\begin{eqnarray}
&&W[\rho,\,S]=-m\,D\,\frac{\partial}{\partial\,x}\,\left(\frac{j_{_{\psi}}}{\rho}\right)
 \ ,\\
&&{\cal
W}[\rho,\,S]=\frac{\hbar\,D}{2\,\rho}\,\frac{\partial^2\,\rho}{\partial\,x^2}
\ .
\end{eqnarray}
The quantum current $j_{_\psi}$ takes the form of a Fokker-Planck current:
\begin{equation}
j_{_{\psi}}=\frac{S_1}{m}\,\rho+D\,\rho_1 \ ,
\end{equation}
resulting to be the sum of two terms, the former is a drift current while the latter
is a Fick current. The generator of the transformation
$\cal U$ (\ref{gauge}) takes the form:
\begin{equation}
{\cal U}[\psi,\,\psi^\ast]=\exp\left(\frac{i}{\hbar}\,m\,D\,\log\rho\right)
\ .\label{tra1}
\end{equation}
This is a particular case of a class of transformations
introduced by Doebner and Goldin and permits to write the evolution
equation:
\begin{equation}
i\,\hbar\,\frac{\partial\,\phi}{\partial\,t}=-\frac{\hbar^2}{2\,m}\,\frac{\partial^2\,\phi}
{\partial\,x^2}+2\,m\,D^2\,\frac{1}{\rho^{1/2}}\,\frac{\partial^2\,\rho^{1/2}}{\partial\,x^2}\,\phi
 \ .\label{27}
\end{equation}
Equation (\ref{27}) was studied in Ref. \cite{Guerra}; after rescaling
the phase: $\sigma\rightarrow\sqrt{1-(2\,m\,D/\hbar)^2}\,\sigma$,
it reduces to the linear Schr\"odinger equation.\\
As a second example we consider a nonlinear Schr\"odinger equation,
introduced in Ref. \cite{Jackiw}, where the complex
nonlinearity is originated by the potential:
\begin{equation}
U[\rho,\,S]=\frac{\hbar^2\,\lambda^2}{8\,m}\,\rho^3-\frac{\hbar\,\lambda}{2\,m}\,\rho^2\,S_1
 \ ,\label{jac}
\end{equation}
and its real and imaginary part take respectively the form:
\begin{eqnarray}
&&W[\rho,\,S]=\frac{3\,\hbar^2\,\lambda^2}{8\,m}\,\rho^2-\frac{\hbar\,\lambda}{m}\,\rho\,S_1\\
&&{\cal W}[\rho,\,S]=\frac{\hbar^2\,\lambda}{2\,m}\,\rho_1 \ .
\end{eqnarray}
Within this model the quantum current is given by:
\begin{equation}
j_{_{\psi}}=\frac{S_1}{m}\,\rho-\frac{\hbar\,\lambda}{2\,m}\,\rho^2 \ .
\end{equation}
The generator $\cal U$ of the transformation (\ref{gauge})
takes the simple form:
\begin{equation}
{\cal U}[\psi,\,\psi^\ast]=\exp\left(-\frac{i\,\lambda}{2}\,\int\rho\,dx\right)
 \ ,
\end{equation}
while the evolution equation for the new complex quantity $\phi$, containing a
real nonlinearity takes the form:
\begin{equation}
i\,\hbar\,\frac{\partial\,\phi}{\partial\,t}=-\frac{\hbar^2}{2\,m}\,\frac{\partial^2\,\phi}
{\partial\,x^2}-\lambda\,\hbar\,j_{_{\phi}}\,\phi \ .
\end{equation}
The last application is given by a nonlinear Schr\"odinger
equation, proposed recently by us \cite{noi1,noi2}, describing a system of particles
obeying to an exclusion-inclusion principle (EIP), which are originated
from a collective interaction.
The potential $U[\rho,\,S]$ in this model is given by:
\begin{equation}
U[\rho,\,S]=\kappa\frac{(S_1\,\rho)^2}{2\,m} \ .\label{eip}
\end{equation}
The real and imaginary part of the nonlinearity in the evolution
equation are:
\begin{eqnarray}
&&W[\rho,\,S]=\kappa\,\frac{m}{\rho}\,\left(\frac{j_{_{\psi}}}{1+\kappa\,\rho}\right)^2
 \ ,\\
&&{\cal
W}[\rho,\,S]=-\kappa\,\frac{\hbar}{2\,\rho}\,\frac{\partial}{\partial\,x}\,
\left(\frac{j_{_{\psi}}\,\rho}{1+\kappa\,\rho}\right) \ .
\end{eqnarray}
while the current $j_{_\psi}$ is given by:
\begin{equation}
j_{_{\psi}}=\frac{S_1}{m}\,\rho\,(1+\kappa\,\rho) \ .
\end{equation}
The parameter $\kappa$ quantifies the EIP, which acts as an
exclusion principle (inclusion) if $\kappa<0$ ($\kappa>0$), since the
factor $1+\kappa\,\rho$, appearing in the expression of $j_{_\psi}$
becomes an inhibition (enhancement) factor. The gauge transformation $\cal
U$ is:
\begin{equation}
{\cal U}[\psi,\,\psi^\ast]=\exp\left(\frac{i\,\kappa}{\hbar}\,\int\rho\,S_1\,dx\right)
 \ ,
\end{equation}
and the evolution equation for the field $\phi$ takes the form:
\begin{eqnarray}
\nonumber
i\,\hbar\,\frac{\partial\,\phi}{\partial\,t}&=&-\frac{\hbar^2}{2\,m}\,\frac{\partial^2\,\phi}
{\partial\,x^2}+\kappa\,m\,\frac{j_{_{\phi}}^2}{\rho\,(1+\kappa\,\rho)}\,\phi\\
&&-\kappa\,\frac{\hbar^2}{4\,m}\,\left[\frac{\partial^2\,\rho}{\partial\,x^2}
-{1\over\rho}\left(\frac{\partial\,\rho}{\partial\,x}\right)^2\right]\,\phi
\ .
\end{eqnarray}
As a conclusion, it is worth to remark that Eq. (\ref{gauge}) defines a wide class of
nonlinear gauge transformations and allows
to reduce the nonlinearity in the evolution equation from complex to
real, in several physical systems.

\vfill\eject
\end{document}